# Quality-Aware Popularity Based Bandwidth Allocation for Scalable Video Broadcast over Wireless Access Networks

Mostafa Zaman Chowdhury[1] and Yeong Min Jang[2]
[1]Dept. of Electrical and Electronic Engineering, Khulna University of Engineering & Technology, Bangladesh
[2]Dept. of Electronics Engineering, Kookmin University, Seoul 136-702, Korea
E-mail: mzceee@yahoo.com, yjang@kookmin.ac.kr

## Abstract

Video broadcast/multicast over wireless access networks is an attractive research issue in the field of wireless communication. With the rapid improvement of various wireless network technologies, it is now possible to provide high quality video transmission over wireless networks. The high quality video streams need higher bandwidth. Hence, during the video transmission through wireless networks, it is very important to make the best utilization of the limited bandwidth. Therefore, when many broadcasting video sessions are active, the bandwidth per video session can be allocated based on popularity of the video sessions (programs). Instead of allocating equal bandwidth to each of them, our proposed scheme allocates bandwidth per broadcasting video session based on popularity of the video program. When the system bandwidth is not sufficient to allocate the demanded bandwidth for all the active video sessions, our proposed scheme efficiently allocates the total system bandwidth among all the scalable active video sessions in such a way that higher bandwidth is allocated to higher popularity one. Using the mathematical and simulation analyses, we show that the proposed scheme maximizes the average user satisfaction level and achieves the best utilization of bandwidth. The simulation results indicate that a large number of subscribers can receive a significantly improved quality of video. To improve the video quality for large number of subscribers, the only tradeoff is that a very few subscribers receive slightly degraded video quality.

## 1. Introduction

Nowadays digital video is the fastest-growing data application and mobile TV has become popular as it promises to deliver video contents to users whenever they want and wherever they are. Mobile TV has already proved to be a very promising ARPU (average revenue per user) generator for cellular operators with several million mobile TV subscribers worldwide [1]. During last couple of years, a notable development of broadband wireless access networks has been observed. Low-cost, high data rate transmission, and better Quality of Service (QoS) provisioning radio technologies are available now. Worldwide Interoperability for Microwave Access (WiMAX) is a typical example of an emerging wireless network system. The emerging Mobile WiMAX (802.16e) is capable of providing high data rate with QoS mechanisms, making the support of mobile TV very attractive [2]. The fast deployment of broadband wireless networks has raised expectation of real-time video services in mobile environments. However, high quality video streaming over wireless networks is challenging issue. Therefore, many works need to be done for efficient deployment of real-time high quality video services over wireless access networks. Transmission of high quality video requires higher amount of bandwidth which is difficult to guarantee because of the resource constraints in wireless networks [3]. A video stream usually requires a few hundreds to a few thousands kbps [4] and the delay-sensitive video transmission requires the efficient handling of wireless link bandwidth. Therefore, transmission of videos through the wireless link using broadcasting or multicasting technique has become very popular approach compared to the unicast approach.

In case of mobile TV, users can enjoy video services anywhere even with full mobility support through the several access networks. Fig. 1 shows an example of basic network architecture for the mobile TV deployment. The mobile TV user may connect with the existing macrocellular networks or Mobile WiMAX networks or femtocell networks or others wireless access networks. However, the quality of video services over wireless access networks is mostly depended on the allocated bandwidth for a broadcasting/multicasting session.

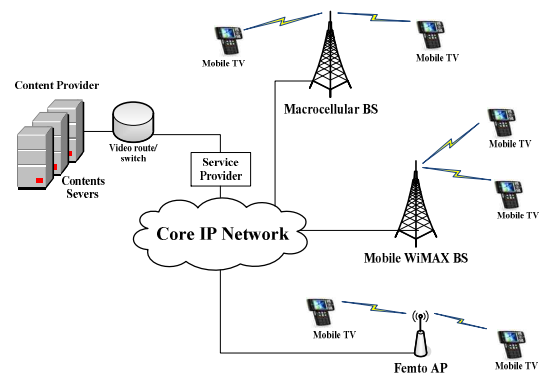

**Figure 1.** Basic network architecture for the video transmission over various wireless networks.

Scalable Video Coding (SVC) is an excellent solution to the problems raised by the diverse characteristics of high data rate video transmission through the wireless link. The SVC allows the elimination of some parts of the video bit stream in order to adapt it to the various needs or preferences of end users as well as to varying terminal capabilities or network conditions [5]. Therefore, scalable video technique [5-8] allows the variable bit rate video broadcast/multicast over wireless networks. This technique utilizes multiple layering. Each of the layers improves spatial, temporal, or visual quality of the rendered video to the user [6]. Base layer or the highest priority layer guarantees the minimum quality of a video stream. The addition of enhanced layers or low priority layers improves the video quality. The number of layers for a video session (program) and the bandwidth per layer can be manipulated dynamically. Thus, to broadcast/multicast videos through a wireless environment, layered transmission is an effective approach for supporting heterogeneous receivers with varying bandwidth requirements [8]. Hence, if the system bandwidth is not sufficient to allocate the demanded bandwidth for all of the active broadcasting/multicasting video sessions, it is possible to allocate higher bandwidth for the popular video session compared to less popular one. In this paper, we address two important problems in video broadcast/multicast over wireless networks: 1) maximizing the average user satisfaction level and 2) the best utilization of the network bandwidth.

The real-time video services require higher bandwidth compared to the other applications such as voice and data. Due to the limited data rates of wireless networks, it is not possible to provide the best quality for the entire active broadcasting/multicasting video sessions. Hence, equal bandwidth allocation for all of the broadcasting/multicasting video sessions is an easy and simple method. The service qualities of all broadcasting/multicasting video sessions are equally degraded when the total wireless bandwidth is not sufficient to provide the maximum demanded bandwidth to all. Instead of allocating equal bandwidths to all of the sessions during an insufficient bandwidth condition, our proposed scheme efficiently allocates the total system bandwidth among them in such a way that higher bandwidth is allocated to the video session of higher popularity. Thus, the average user satisfaction level is increased significantly. However, a minimum quality for the lowest popular video session is guaranteed by assigning a minimum amount of bandwidth.

This paper is organized as follows. Section 2 provides the background and related works for the video delivery over wireless link. The proposed bandwidth allocation scheme is presented in Section 3. Section 4 provides the performance analysis using the simulation results for the proposed scheme. Finally, we give our conclusion in Section 5.

## 2. Background and Related Works

### 2.1 Background

The video streaming over wireless networks is composed of three main entities, namely content source, base station, and subscribers [6]. Video contents are aggregated from different sources and sent to the base station. The base station transmits the incoming videos to the subscribers. Most of the existing wireless network technologies such as femtocell [9], WiFi, Mobile WiMAX, 3G, and 4G support multicast/broadcast mechanisms [6, 10-13]. Especially the features such as, high data rate, flexible QoS mechanisms, multicast/broadcast, and delivery of multimedia content to large-scale user communities in cost-efficient manner of Mobile WiMAX (802.16e) network make it as a pioneer wireless technology for the provisioning of video streams over wireless networks. UDCast [1] is going to develop broadcast TV services supporting around 50 channels over Mobile WiMAX.

Bandwidth is a limited resource in wireless networks to support high data rate video services. Wireless networks can support a fraction of wired networks. For example the 10 Gigabit Ethernet can support maximum 10 Gbps, while the 4G Mobile WiMAX networks can support maximum 100 Mbps. Even future 5G networks have to support very high data rate video services. Hence, the supporting of video services through the wireless networks is still challenging and proper management of the resources is essential.

Three type of approaches are used for the video streaming over wireless networks: unicast, multicast, and broadcast [10]. The unicast approach is usually employed by on-demand streaming. However, the multicast and broadcast approaches [14] are often employed by live streaming. Since bandwidth is a scarce resource for the wireless networks, it is efficient to use multicast/broadcast to deliver video services over wireless networks. Normally, VoD (video on demand) systems can be categorized into True-VoD (TVoD), which is based on unicast transmission, and Near-VoD (NVoD), which is based on broadcast or multicast transmission [15]. The NVoD methods can be classified into three main approaches, batching, patching, and broadcasting. The batching approach collects set requests that arrives in close time, and then serves all of them together with one channel. In patching, video request is firstly served by unicast stream and then joined back to a multicast stream. In broadcasting approach, the video is periodically broadcast into dedicated channel with pre-defined schedule [15]. When many users simultaneously want to watch the same video program (e.g., movie or live sports), wireless-link bandwidth would fall short if a separate point-to-point channel (unicast approach) is required for each user [11]. Therefore, the broadcasting and multicasting approaches have become very popular to handle the limited wireless link bandwidth. In broadcasting and multicasting approaches, video sessions are efficiently delivered to many users in parallel. Wireless broadcasting and multicasting techniques enable various mobile users with different platforms to access to the multimedia information simultaneously [7].

Another issue is the SVC technique for the layered video streaming over wireless networks. It gives simple solutions for adaptation to network constraint and user capabilities by providing a full scalability including temporal, spatial, and quality scalability [7]. SVC enables the transmission and decoding of partial bit streams to provide video services with lower temporal or spatial resolutions or reduced fidelity while retaining a reconstruction quality that is high relative to the rate of the partial bit streams [5]. There are three approaches of scalability, namely, temporal, spatial, and quality scalability. The substream of the spatial scalability provides the source content with a reduced picture size (spatial resolution). The substream of the temporal scalability provides the source content with a reduced frame rate (temporal resolution). Finally, the substream of the quality scalability provides the same spatio–temporal resolution as the complete bit stream, but with a lower fidelity (signal-to-noise ratio). A scalable video stream [5-8] is composed of multiple layers, where each layer improves the spatial, temporal, or the visual quality of the rendered video to the user [6]. Thus, the scalable video coding is an attractive solution for the multi-rate video transmission over wireless networks. It encodes raw video data into several layers with different priority. The highest priority layer is called the base layer. The base layer contains the data with the highest importance, which can provide a minimum video quality. Some additional lower priority layers, called enhanced layers, contain data that progressively enhance the reconstructed video quality of the base layer [7, 8]. The layers are then distributed to receivers by a layered transport protocol. Both the bandwidth of a layer and the number of layers can be dynamically manipulated with a fast response time [8]. Thus, the video resolution and quality that can be offered to customers mainly depend on the bit rate allocated to the video session.

## 2.2 Related Works

In the past few years, there has been extensive works on video broadcast/multicast over wireless networks [2-8, 10, 11, 14, 16-22]. Jianfeng Wang et al. [2] proposed an end-to-end solution for the multicast/broadcast service (MBS) based on cross-layer optimization. The scheme addressed the synchronization, energy efficiency, and robust video quality issues. Juan Carlos Fernandez et al. [3] proposed a dynamic QoS negotiation scheme. Their proposed algorithm allows the users to dynamically negotiate the service levels required for their traffic and to reach them through one or more wireless interfaces. The proposed QoS negotiation system ensures the continuity of QoS perceived by mobile users while they are on the move between different access points. Jen-Wen Ding et al. [4] proposed a bandwidth allocation scheme that can dynamically adjust the bit rates allocated to different videos. Heiko Schwarz et al. [5] presented an overview of of the basic concepts for extending H.264/AVC towards SVC. Moreover, the basic tools for providing temporal, spatial, and quality scalability are also described in detail. Somsubhra Sharangi et al. [6] presented the energy-efficient multicasting of scalable video streams over WiMAX networks. Yu Wang et al. [7] presented a variable bit rate allocation for the broadcasting of scalable video over wireless networks. They proposed variable bit rate allocation for the base layer as well as for the enhanced layers. Jiangchuan Liu et al. [8] proposed the layering in multisession video broadcasting. The approach employed a generic utility function for each receiver under each video session. Jen-Wen Ding et al. [10] proposed the spectrum-based bandwidth allocation algorithm for layered video streams over wireless broadcast channels. Their proposed algorithm uses the weighted spectrum to reflect the importance of the quality of different video sessions. Stream rate adapter is used for the dynamically adjusting the bit rate of layered video streams according to the available bandwidth in wireless access networks. Ji Hoon Lee et al. [11] presented an MBS architecture that is based on location-management areas. Zeng-Yuan Yang et al. [16] presented a broadcasting scheme that shows the relationship between the delay and the minimum bandwidth requirement. Stuart Pekowsky et al. [17] presented an overview of multimedia data broadcasting strategies.

Our proposed scheme allocates the bandwidth per broadcasting video session based on the popularity of the video session. Allocation of higher bandwidth means, the broadcasting/multicasting video session may contains more number of enhanced layers or each of the enhanced layers may have higher bandwidth to improve the video quality.

## 3. Proposed Bandwidth Allocation Scheme

Even though the effective bandwidth of wireless links is growing very rapidly, fully deployed 4G wireless network will not even enough to accommodate many best quality video services simultaneously. The wireless link will always have less bandwidth than the wired links and it will continue in further. Efficient bandwidth allocation for the video sessions in broadcasting is needed to make the best usage of the scare resources of wireless networks. An easy and straightforward approach is that all of the active broadcasting video sessions share the total system bandwidth equally. However, such approach is not sensible. Because a popular video program attracting a large number of subscribers should be allocated with more bandwidth compared to the less popular one, if allocation of total demanded bandwidth is not possible. Therefore, we propose popularity based efficient bandwidth allocation scheme that makes the best utilization of the bandwidth. The notations used in this section for different equations are summarized in Table 1.

**Table 1** Notations for the analysis

| Notation | Explanation |
|---|---|
| $C$ | The total system bandwidth capacity to support scalable broadcasting video services |
| $\beta_{max}$ | Maximum allocated bandwidth for each of the broadcasting sessions |
| $\beta_{min}$ | Minimum allocated bandwidth for each of the broadcasting sessions |
| $K$ | Total number of active users in the system |
| $K_m$ | Number of users watching the $m$-th broadcasting video program (session). $m=1$ indicates that video program which is watched by maximum number of users. $m=M$ indicates that video program which is watched by minimum number of users. $K_1 \geq K_2 \geq \cdots \geq K_m \geq \cdots \geq K_M$ |
| $M$ | Number of active broadcasting sessions |
| $N_{HQ}$ | Minimum number of video sessions that can be provided simultaneously with the allocated bandwidth of $\beta_{max}$ (best quality) for each of them. |
| $N_{LQ}$ | Maximum number of video sessions than can be provided simultaneously by the system with the allocated bandwidth $\beta_{min}$ (lowest quality) for each of them. |
| $\beta$ | Allocated bandwidth for each of the broadcasting video sessions by the equally shared bandwidth allocation scheme |
| $S_L$ | User satisfaction level for the equally shared bandwidth scheme |
| $\beta_m$ | Allocated bandwidth for the $m$-th video session in the proposed scheme |
| $S_{L(m)}$ | User satisfaction level of the users who are watching the program of the $m$-th broadcasting video session |

Let the total system bandwidth capacity to support scalable broadcasting video services and the total number of active broadcasting video sessions are $C$ and $M$, respectively. $\beta_{max}$ and $\beta_{min}$ are, respectively, the maximum allocated bandwidth and the minimum allocated bandwidth for each of the active broadcasting/multicasting video sessions. Then the allocated bandwidth for each of the active sessions in the equally shared bandwidth allocation scheme is:

$$\beta = \begin{cases} \beta_{max}, & \beta_{max}M \leq C \\ \dfrac{C}{M}, & \beta_{max}M > C \end{cases} \quad (1)$$

As we mentioned before, larger amount of allocated bandwidth to a broadcasting video session makes the chance of increasing the number of enhanced layers and thus improving the video quality for that session. User satisfaction level depends on received video quality. Therefore, we assume that the user satisfaction level is directly proportional to allocated bandwidth for a broadcasting video session. User satisfaction level becomes maximum (equal to 1) when the demanded bandwidth ($\beta_{max}$) is allocated for a broadcasting video session. The satisfaction level of a user in the equally shared bandwidth allocation scheme can be written as:

$$S_L = \frac{\beta}{\beta_{max}} = \begin{cases} 1, & \beta_{max}M \leq C \\ \dfrac{C}{\beta_{max}M}, & \beta_{max}M > C \end{cases} \quad (2)$$

Our proposed scheme allocates different amount of bandwidths for different broadcasting video sessions based on popularity of video programs. However, the maximum allocated bandwidth to a broadcasting video session is $\beta_{max}$ and the minimum allocated bandwidth to a broadcasting video session is $\beta_{min}$. This $\beta_{min}$ amount of bandwidth ensures the minimum quality of a video session. An active broadcasting video session is ranked based on the number of users currently watching the program on that session. The most popular broadcasting video session (program) is ranked as 1 whereas the lowest popular one is ranked as $M$. The numbers of active users for different broadcasting/multicasting video sessions are related as:

$$K_1 \geq K_2 \geq \cdots \geq K_m \geq \cdots \geq K_M \quad (3)$$

where $K_m$ is the number of users watching the $m$-th broadcasting video program. $m=1$ indicates that program which is being watched by the maximum number of users. Whereas $m=M$ indicates that with the minimum users. $K$ is the total number of active users in the system.

Total number of active users, $K$, in the system is:

$$K = K_1 + K_2 + \cdots + K_m + \cdots + K_M \quad (4)$$

Equation (3) can be rewritten as:

$$K_1 \geq \frac{K}{M} \text{ and } K_M \leq \frac{K}{M} \quad (5)$$

Minimum number of broadcasting video sessions, $N_{HQ}$, that can be provided simultaneously with the allocated bandwidth $\beta_{max}$ (best quality) for each of the broadcasting/multicasting video sessions is:

$$N_{HQ} = \left\lfloor \frac{C}{\beta_{max}} \right\rfloor \quad (6)$$

Maximum number of broadcasting video sessions, $N_{LQ}$, that can be provided simultaneously with the allocated bandwidth $\beta_{min}$ (lowest quality of video) for each of the broadcasting video sessions is:

$$N_{LQ} = \left\lfloor \frac{C}{\beta_{min}} \right\rfloor \quad (7)$$

Fig. 2 shows the basic concepts of bandwidth allocation per broadcasting video session by the equally shared and the proposed popularity based bandwidth allocation schemes when the system bandwidth is not sufficient to allocate $\beta_{max}$ for each of the active broadcasting video sessions. Fig. 2(a) shows that an equal bandwidth $\beta$ is allocated to each of the broadcasting video sessions by the equally shared bandwidth allocation scheme. On the other hand, Fig. 2(b) shows that the same bandwidth is not allocated to each of the active broadcasting video sessions by the proposed popularity based bandwidth allocation scheme. Maximum bandwidth $\beta_1$ is allocated to the broadcasting casting video session #1 which is enjoyed by the maximum number of subscribers. On the other hand, minimum bandwidth $\beta_M$ is allocated to the broadcasting video session #$M$ which is received by the minimum number of subscribers.

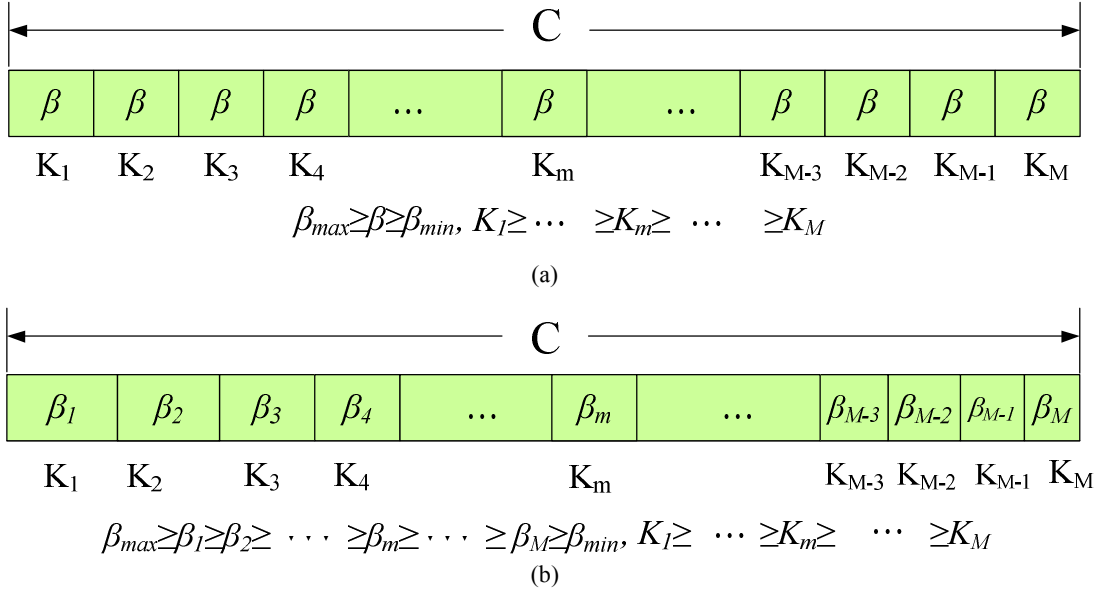

**Figure 2.** An example of bandwidth allocation when the system bandwidth is not sufficient to allocate $\beta_{max}$ for each of the broadcasting video sessions (a) equal allocated bandwidths to all of the broadcasting video sessions by the equally shared bandwidth allocation scheme, (b) different allocated bandwidths to different broadcasting video sessions by the proposed popularity based bandwidth allocation scheme.

Bandwidth $\beta_{max}$ is allocated for each of the broadcasting video sessions whenever $\beta_{max}M \leq C$. However, if $\beta_{max}M > C$, then the allocated bandwidth $\beta_m$ for $m$-th broadcasting video session in the proposed popularity based bandwidth allocation scheme is calculated by the following procedures,

$$X_m = \begin{cases} 0, & \left(aK_m + \sum_{j=1}^{m-1} X_j\right) \leq \beta_{diff} \\ \dfrac{aK_m + \sum_{j=1}^{m-1} X_j - \beta_{diff}}{M - m}, & \left(aK_m + \sum_{j=1}^{m-1} X_j\right) > \beta_{diff} \end{cases} \quad (8)$$

where $a = \dfrac{M}{K}\left(\dfrac{C}{M} - \beta_{min}\right)$ and $\beta_{diff} = \beta_{max} - \beta_{min}$

$$\beta_m = \begin{cases} \beta_{max}, & \left(aK_m + \sum_{j=1}^{m-1} X_j\right) \geq \beta_{diff} \\ \beta_{min} + aK_m + \sum_{j=1}^{m-1} X_j, & \left(aK_m + \sum_{j=1}^{m-1} X_j\right) < \beta_{diff} \end{cases} \quad (9)$$

Hence, the dedicated bandwidth for the $m$-th broadcasting/multicasting video session is $\beta_m$. However, for instance, a receiving device with limited resources e.g., restricted display resolution, processing capacity, and battery power decodes only a part of the broadcasted/multicasted bit stream. Therefore, in a broadcast/multicast cases, terminals with different capabilities can be served through a single scalable bit stream.

The difference between the allocated bandwidth for the $m$-th and the $(m+1)$-th broadcasting video session is:

$$\beta_m - \beta_{m+1} = \begin{cases} 0, & \left(aK_m + \sum_{j=1}^{m+1} X_j\right) \geq \beta_{diff} \text{ or} \\ & \left(K_m = K_{m+1}\right) \text{ or } \dfrac{C}{m} \geq \beta_{max} \\ a(K_m - K_{m+1}) - X_m, & \left(aK_m + \sum_{j=1}^{m+1} X_j\right) < \beta_{diff} \end{cases} \quad (10)$$

Hence, from (8)-(10), the allocated bandwidths of the active broadcasting/multicasting video sessions for the proposed scheme are related as follows,

Case 1: when $\beta_{max}M \leq C$,

$$\beta_1 = \beta_2 = \cdots = \beta_m = \cdots = \beta_M = \beta_{max} \quad (11)$$

Case 2: when $\beta_{max}M > C$,

$$\left. \begin{array}{l} \beta_1 \geq \beta_2 \geq \cdots \geq \beta_m \geq \cdots \geq \beta_M \\ \beta_{max} \geq \beta_1 \geq \dfrac{C}{M} \\ \beta_{min} \leq \beta_M \leq \dfrac{C}{M} \end{array} \right\} \quad (12)$$

Whenever the number of users for an active broadcasting video session or the total number of active broadcasting video sessions is changed, the allocated bandwidth for each of the active broadcasting/multicasting video sessions is also dynamically changed. As a consequence, the number of enhanced layers per session and the allocated bandwidth per enhanced layer may also be changed. It can be mentioned that a receiver cannot subscribe to a fraction of a layer.

In our proposed scheme, the satisfaction level of the users who are connected with the *m-th* broadcasting/multicasting video session is:

$$S_{L(m)} = \begin{cases} 1, & \beta_{max}M \leq C \\ \dfrac{\beta_m}{\beta_{max}}, & \beta_{max}M > C \end{cases} \quad (13)$$

From (11)-(13), the relation between the satisfaction levels of different users can be written as:

$$1 \geq S_{L(1)} \geq S_{L(2)} \geq \cdots \geq S_{L(m)} \geq \cdots \geq S_{L(M)} \geq \dfrac{\beta_{min}}{\beta_{max}} \quad (14)$$

The average user satisfaction level for the proposed scheme is calculated as:

$$S_{L(av)} = \begin{cases} 1, & \beta_{max}M \leq C \\ \dfrac{\sum_{m=1}^{M} S_{L(m)}K_m}{K}, & \beta_{max}M > C \end{cases} \quad (15)$$

where $S_{L(av)}$ is the average user satisfaction level for the proposed scheme considering all the active users in the system.

The relation between the average user satisfaction levels for the proposed popularity based bandwidth allocation scheme and the equally shared bandwidth allocation scheme can be written as:

$$\left.\begin{aligned} S_{L(av)} &= S_L = 1, & \beta_{max}M \leq C \\ S_{L(av)} &= S_L, & K_1 = K_M \\ S_{L(av)} &> S_L, & K_1 \neq K_M \text{ and } \beta_{max}M > C \end{aligned}\right\} \quad (16)$$

It seems that, if larger number of users watch the program of a broadcasting video session, then higher bandwidth is allocated for that video session to provide better quality of service for those. Thus, the average user satisfaction level is increased significantly.

## 4. Performance Evaluation

In this section, we verified performance of the proposed scheme using simulation results. The basic assumptions for the performance analysis are shown in Table 2. We performed the analysis for two scenarios of traffic environment while the total number of users in the system is fixed. Scenario 1 considers random number of active users for per video session. Scenario 2 also considers random manner. However, in second scenario,

**Table 2** Basic assumptions

| Parameter | Value |
|---|---|
| The total system bandwidth capacity *(C)* | 30 Mbps |
| Maximum allocated bandwidth for each of the broadcasting/multicasting video sessions ($\beta_{max}$) | 2 Mbps |
| Minimum allocated bandwidth for each of the broadcasting/multicasting video sessions ($\beta_{min}$) | 0.6 Mbps |
| Number of users with per video session | Random |
| Total number of active users in the system (*K*) | 200 |

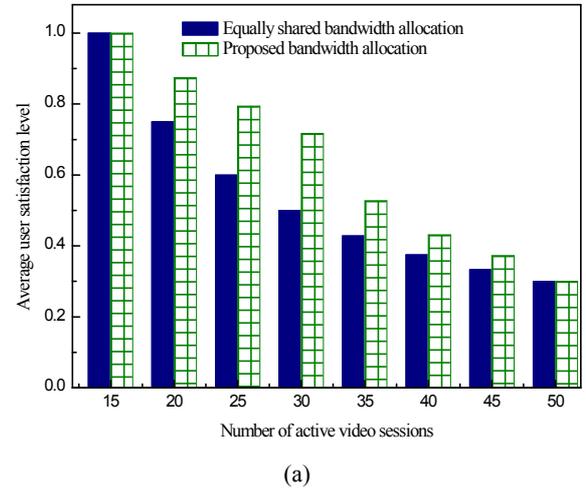

(a)

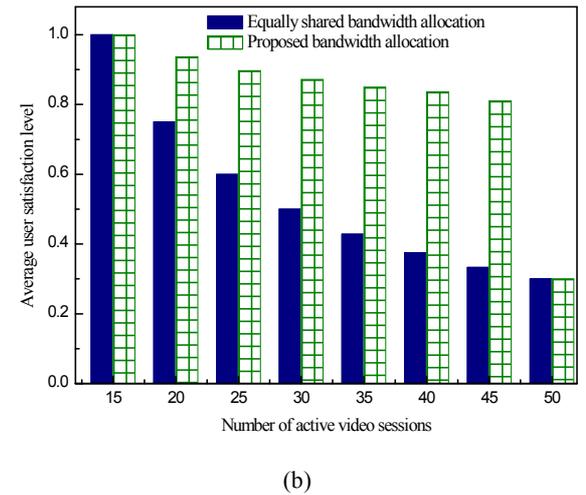

(b)

**Figure 3.** A comparison of the average user satisfaction levels for various numbers of active video sessions (a) scenario 1 traffic environment, (b) scenario 2 traffic environment.

50% users watch one video program and the remaining 50% users watch other video programs.

Firstly, we verify the improvement of average user satisfaction level for our proposed scheme compared to the equally shared bandwidth allocation scheme. Fig. 3 shows that the proposed scheme provides much better average user satisfaction level compared to the equally shared bandwidth allocation scheme. The user satisfaction level decreases with the increase of active video sessions due to the limited bandwidth capacity of the network. Fig. 3(b) shows that our proposed scheme is even more effective when large number of users watch the program of a common broadcasting/multicasting video session.

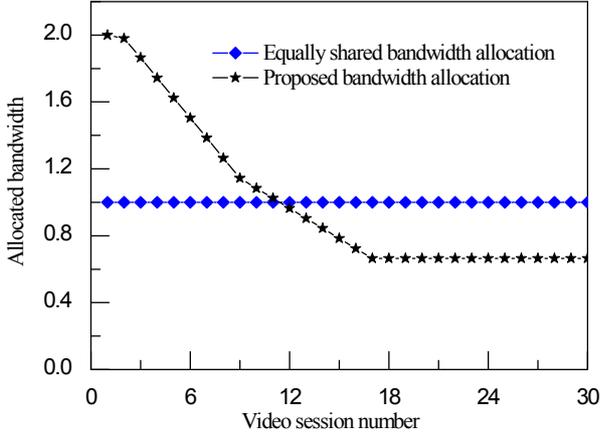

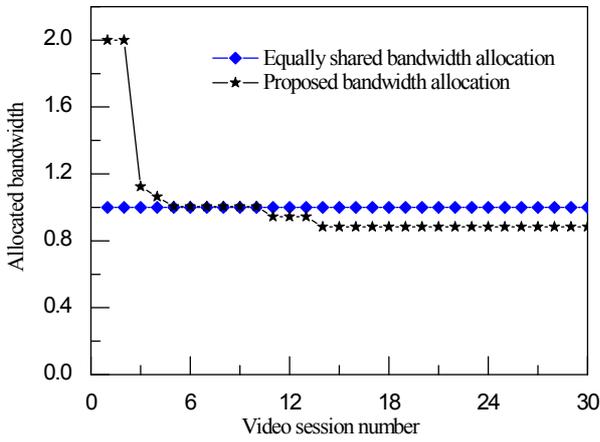

**Figure 4.** Allocated bandwidths for various video sessions when 30 video sessions are active (a) scenario 1 traffic environment, (b) scenario 2 traffic environment.

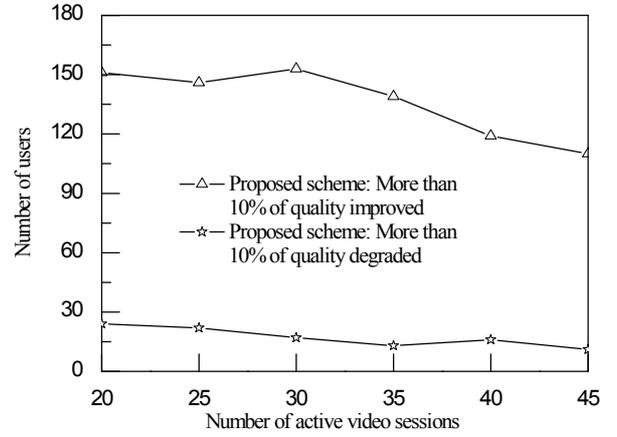

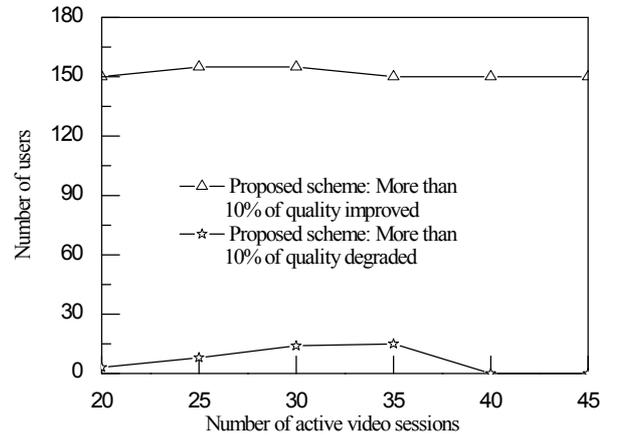

**Figure 5.** A Comparison to show the number of users to whom video quality is improved or degraded with respect to the equally shared bandwidth allocation scheme (a) scenario 1 traffic environment, (b) scenario 2 traffic environment.

Fig. 4 shows the allocated bandwidth for each of the video sessions when 30 video sessions are active. It shows that the allocated bandwidth to a video session is gradually decreased with the decrease of popularity in the proposed scheme. The maximum allowable bandwidth $\beta_{max}$ can be allocated for more than one video sessions depending on the network bandwidth and the traffic conditions. However, the allocated bandwidth for any of the active broadcasting/multicasting video sessions does not go below a threshold level to guarantee the minimum video quality for all the active video sessions. Hence, the allocated bandwidths for some broadcasting/multicasting video sessions are increased and for some broadcasting/multicasting video sessions are decreased compared to the equally shared bandwidth scheme. In Fig. 4(a) case, 168 users enjoy improved quality videos whereas 32 users receive slightly degraded quality of videos. In Fig. 4(b) case, 177 users enjoy improved quality videos whereas 23 users receive slightly degraded quality of videos. However, for both the cases minimum quality of video is assured.

Fig. 5 shows a comparison between the numbers of users to whom the video quality is improved and the users to whom it is degraded in the proposed scheme compared to the equally shared bandwidth allocation scheme. Both Figs. in 5(a) and 5(b) indicate that huge number of users can enjoy improved video quality. To improve the video quality for these large number of users, the only adjustment is that a very few users receive slightly degraded video quality. Hence, a large number of users enjoy the significantly improved video quality.

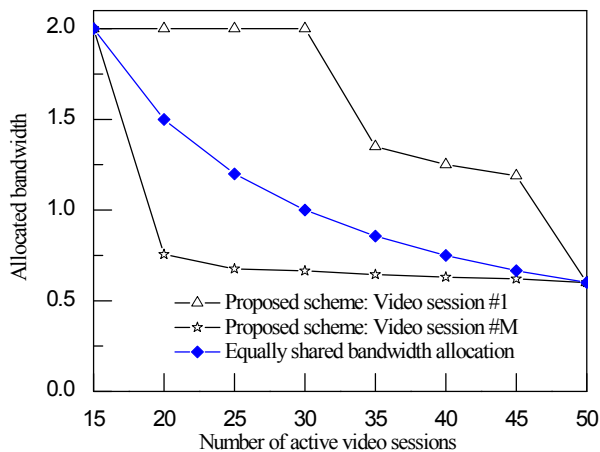

(a)

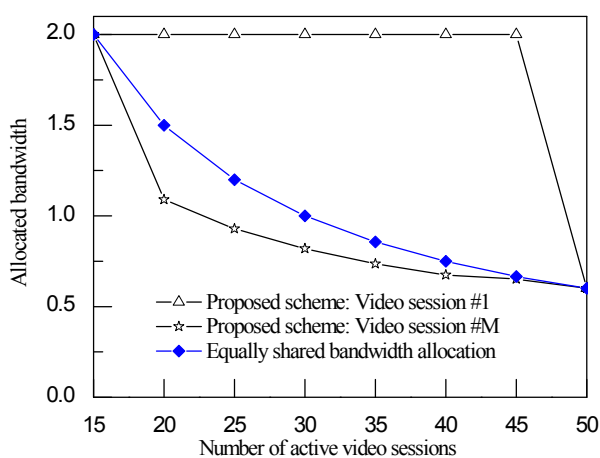

(b)

**Figure 6.** A comparison of the allocated bandwidths for the most popular video session and the lowest popular video session (a) scenario 1 traffic environment, (b) scenario 2 traffic environment.

Fig. 6 demonstrates the bandwidth allocations for the most popular broadcasting/multicasting video session and lowest popular broadcasting/multicasting video session in the proposed scheme. Both Figs. 6(a) and 6(b) designate that the most popular video session is guaranteed with the sufficient bandwidth except the case where entire system bandwidth is required to provide the minimum bandwidth for each of the video sessions. For the Fig. 6(b), even the bandwidth allocation for the lowest popular video session is quite close to that of the equally shared bandwidth allocation scheme.

The results of the performance analyses show that our proposed popularity based bandwidth allocation scheme is able to improve average user satisfaction level within the limited bandwidth. The proposed scheme is even more effective when large number of users watch the program of a common broadcasting/multicasting video session.

## 5. Conclusions

It is expected that the next generation wireless networks will fully support the high quality real-time video services. However, management of the scare wireless bandwidth is a challenging issue for supporting the high quality video services through wireless access networks. This paper proposes an efficient bandwidth allocation scheme for the real-time video broadcast/multicast over wireless networks. The proposed scheme allocates bandwidth for each of the broadcasting/multicasting video sessions based on the importance of the sessions during the lack of bandwidth situation. The entire active video sessions are ranked based on the number of users watching the program of a video session. The allocated bandwidth for each of the video sessions is dynamically changed with the changing numbers of connected users to different video sessions or with the change of number of active video sessions. We compare the proposed scheme with the equally shared bandwidth allocation scheme to show the performance improvement. This paper also demonstrates how the popularity of a video session affects the bandwidth allocation. Simulation results indicate that the proposed bandwidth allocation scheme is very effective for video broadcast/multicast over the wireless networks.

## Acknowledgement

This research was supported by Basic Science Research Program through the National Research Foundation of Korea (NRF) funded by the Ministry of Education (no.NRF-2013R1A1A2057922).

## Biographies

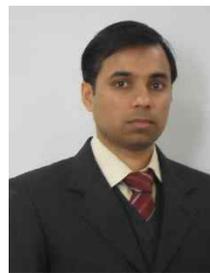

**Mostafa Zaman Chowdhury** received his B.Sc. degree in electrical and electronic engineering from Khulna University of Engineering and Technology (KUET), Bangladesh, in 2002. He received his M.Sc. and Ph.D. degrees both in electronics engineering from Kookmin University, Korea, in 2008 and 2012, respectively. Currently he is working as an Associate Professor at KUET, Bangladesh. His research interests include 5G mobile communications, convergence networks, and QoS provisioning.

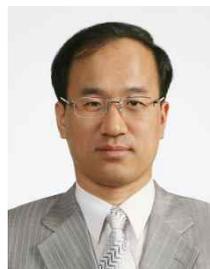

**Yeong Min Jang** received the B.E. and M.E. degrees both in electronics engineering from Kyungpook National University, Korea, in 1985 and 1987, respectively. He received the doctoral degree in computer science from the University of Massachusetts, USA, in 1999. Since 2002, he is with the School of Electrical Engineering, Kookmin University, Korea. His research interests include 5G mobile communications, multi-screen convergence, public safety, optical wireless communications, optical camera communication (OCC), and internet of things (IoT).